# Unraveling the Redox Mechanisms Underlying FLASH Radiotherapy: Critical Dose Thresholds and NRF2-Driven Tissue Sparing

Yan Zhang 1†, Chenyang Huang 1†, Ankang Hu2,5, Yucheng Wang1, Yixun Zhu1, Wanyi Zhou 2,5, Jiaqi Qiu6, Jian Wang6, Qibin Fu1,4* Tuchen Huang 1,3,4* Hao Zha 2,5*, Wei Wang1,3,4, Junli Li2,5

1 Sino-French Institute of Nuclear Engineering and Technology, Sun Yat-sen University, Zhuhai, 519082, China
2 Department of Engineering Physics, Tsinghua University, Beijing, 100084, China
3 GuangDong Engineering Technology Research Center of Nuclear Safety and Emergency Technology, Zhuhai 519082, China
4 Research Center for Nuclear Technology and Applications, Sun Yat-sen University, Zhuhai 519082, China
5 Key Laboratory of Particle & Radiation Imaging, Tsinghua University, Ministry of Education, Beijing, 100084, China
6 ThaccEV Company Limited, Beijing, 100084, China
†These authors contributed equally to this work.

*Correspondence:
Qibin Fu: fuqibin@mail.sysu.edu.cn
Tuchen Huang: huangtuchen@mail.sysu.edu.cn
Hao Zha: zha_hao@mail.tsinghua.edu.cn

**Funding:** This study was supported by the National Key Research and Development Program of China (Grant No. 2022YFC2402300) and the National Natural Science Foundation of China (Grant No. 12441518, 12422511).

Keywords: FLASH radiotherapy, Normal tissue sparing, Dose threshold, Redox mechanisms, NRF2 signaling

## Abstract

FLASH radiotherapy (FLASH-RT) achieves tumor control comparable to conventional dose-rate irradiation (CONV-RT) while significantly reducing radiation damage to normal tissues. However, the physical conditions triggering the FLASH sparing effect remain unclear, and mechanisms related to oxidative stress and redox regulation are poorly understood. This study utilizes a murine acute intestinal toxicity model to investigate how beam parameters influence the FLASH sparing effect and tumor control using innovative FLASH-RT and CONV-RT combined irradiation. Results demonstrate for the first time that a substantially reduced FLASH dose can still elicit sparing effect, provided a total dose threshold is met. Kinetic simulation and experimental validation demonstrate that FLASH-RT enhances peroxyl radical recombination, reducing reactive oxygen species (ROS) and malondialdehyde levels. Antioxidant interventions further confirm the essential role of free radicals. RNA sequencing and molecular analyses reveal that FLASH-RT activates the nuclear factor E2-related factor 2 (NRF2) antioxidant pathway while suppressing extracellular signal-regulated kinases (ERK) signaling, thereby enhancing cellular redox defenses, reducing apoptosis, and mitigating ROS-mediated tissue injury. These findings highlight the feasibility of optimizing the FLASH-RT therapeutic window through redox modulation and provide a foundation for developing free radical-targeted strategies to improve its therapeutic efficacy.

## 1. Introduction

FLASH radiotherapy (FLASH-RT), characterized by a dose delivery rate several orders of magnitude higher than conventional radiotherapy (CONV-RT), maintains comparable tumor control capabilities to CONV-RT while significantly mitigating radiation damage to normal tissues, a phenomenon termed the FLASH effect. FLASH effect has been observed in different subjects and biological endpoints with various types of radiation.[1-12] It is generally believed that a high mean dose rate (MDR > 40 Gy/s) is the key trigger for the FLASH effect. Through a systematic analysis of existing experimental data, we have quantitatively determined the MDR and dose thresholds required to trigger the FLASH effect in the brain and small intestine.[13] However, some studies have revealed that factors such as pulse dose can also influence the FLASH effect. For instance, Liu et al. reported that when the pulse dose reached as high as 4.7 Gy/pulse, the FLASH effect could still be detected even at an MDR as low as 0.3 Gy/s.[7] But recent findings by Veljko et al. contend that the MDR constitutes the principal determinant of the FLASH effect, showing that an MDR exceeding 100Gy/s is required to observe the FLASH effect for small intestine irradiated with 17 Gy.[14] Many previous studies failed to provide complete beam parameter information, hindering the comparative analysis of other contributing factors. Additionally, for pulsed beams, the irradiation time spans vary significantly across different experiments, making simple comparisons based on MDR (calculated as total dose divided by total time) inadequate. Therefore, it is of significant importance to further clarify the conditions required to induce the FLASH effect and elucidate the influences of physical factors such as total dose, MDR and pulse dose. This is particularly critical for beams with pulsed structures, such as electron FLASH irradiation systems.

FLASH irradiation is typically completed within 1 second. The radiochemical stage following ionization plays a critical role in determining the ultimate biological effects, encompassing both the heterogeneous phase of reactive oxygen species (ROS) formation (occurring on nanosecond to microsecond timescales) and the homogeneous phase of ROS interactions (progressing from microseconds to milliseconds).[15-17] This is particularly important for low linear energy transfer (LET) radiation (e.g., X-rays and electrons) since damage is primarily induced through ROS-mediated indirect actions. ROS encompass both radical and non-radical derivatives. The non-radical derivatives include hydrogen peroxide ($H_2O_2$), organic hydroperoxides (ROOH), superoxide anion radical ($O_2^-$), hydroxyl radical (OH•), peroxyl radical (ROO•), etc.[18] These ROS can induce oxidative damage to cellular biomolecules (such as proteins, lipids, and DNA), consequently triggering oxidative stress. The free radical recombination hypothesis proposed in recent years suggests that FLASH irradiation generates ultra-high transient concentrations of peroxyl radicals, thereby increasing the probability of radical recombination.[19] These recombination reactions yield non-radical products, consequently reducing radical-mediated oxidative damage. The subsequently proposed free radical recombination-antioxidant hypothesis introduces competitive reactions between antioxidants and peroxyl radicals,[20] providing further explanation for the differential responses of tumor versus normal tissues to FLASH irradiation. Therefore, elucidating the differences in the response to oxidative stress between FLASH-RT and CONV-RT is crucial for understanding the mechanism underlying the sparing effect of FLASH-RT.

In addition to generating free radicals and inducing oxidative stress, radiation can also perturb intracellular redox homeostasis. These stress responses are regulated by transcription factors such as NRF2, nuclear factor- κB (NF-κB), hypoxia- inducible factor 1 (HIF-1), and heat shock factor 1 (HSF-1). Among them, NRF2 plays a pivotal role in mitigating oxidative stress and maintaining cellular adaptability. By binding to antioxidant response elements (AREs) in the promoter regions of target genes, NRF2 upregulates antioxidants and phase II detoxifying enzymes, such as NAD(P) H: quinone oxidoreductase 1 (NQO1), glutathione-S-transferase (GST), and haem oxygenase 1 (HO-1).[21,22] Recent studies have revealed that the regulatory influence of NRF2 has expanded to encompass proteostasis, metabolic regulation, and iron homeostasis, which are critical for cellular adaptation to environmental and metabolic stresses.[23-25] Furthermore, multiple kinases (e.g., PI3K, AKT, ERK) and transcription factors that induce cell proliferation and differentiation are also redox-regulated.[26] Notably, ERK can participate in pro-apoptotic signaling under specific conditions, such as sustained high-intensity activation or exacerbated oxidative stress.[27] However, the response characteristics of FLASH-RT to oxidative stress and the underlying redox regulatory mechanisms remain poorly understood.

To address these questions, this study employed an innovative approach combining FLASH-RT and CONV-RT irradiation using electron FLASH beams and a murine acute intestinal toxicity model to systematically evaluate the influence of different beam parameters on FLASH sparing effect and the tumor control efficacy of FLASH-RT. The relationships between free radical recombination and radiation dose/dose rate parameters were quantitatively modeled by simulation, with experimental validation through antioxidant intervention to elucidate the role of radicals in mediating the FLASH effect. Furthermore, RNA sequencing (RNA-seq) was employed to analyze and compare gene expression profiles and enrichment pathways in mice under various irradiation conditions, with and without antioxidant treatment. Finally, the changes of NRF2 and its downstream genes, as well as the ERK pathway, were investigated at both the cellular and tissue level to explore the mechanisms underlying the FLASH sparing effect. The findings of this study demonstrate for the first time that FLASH-RT dose can be substantially reduced while still triggering a sparing effect, provided the total dose reaches a threshold level. Moreover, FLASH-RT exhibits tumor control efficacy comparable to that of CONV-RT. FLASH-RT confers tissue sparing by activating the NRF2 antioxidant pathway while suppressing MAPK signaling, thereby enhancing cellular redox defenses and attenuating apoptosis, which ultimately reduces ROS-mediated tissue damage. These findings provide novel perspectives for both mechanistic exploration and clinical translation of FLASH-RT.

## 2. Results

### 2.1. Beam Parameters Influencing the FLASH Effect

To systematically investigate the key factors modulating the FLASH effect, consistency was maintained in mouse strain, sex and age across all experimental groups. Given the peak severity of acute intestinal damage occurring at 3–4 days post-irradiation, small intestine histopathology was evaluated at the 72-hour time point. Due to potential variability in endpoints, this study employed a comprehensive histological scoring system previously described in the literature (Table S1, Supporting information).[28-31] Consistent conclusions were obtained when the quantity of crypt was adopted as the biological endpoint (Figure S1, Supporting information). Table 1 summarizes the irradiation beam parameters for all experimental groups.

As shown in **Figure 1**A, intestinal damage was compared between CONV-RT and FLASH-RT across doses ranging from 6 to 15 Gy. A clear dose-dependent exacerbation of intestinal injury was observed, characterized by inflammatory cell infiltration, disruption of crypt architectural integrity and diminished crypt regenerative capacity. Univariate analysis was employed to systematically evaluate the determinants of the FLASH effect. First, FLASH-RT was delivered at an MDR of 900 Gy/s across varying doses. As shown in Figure 1B, a significant reduction in intestinal damage compared to CONV-RT was only observed at doses reaching 15 Gy, confirming the existence of a minimal total dose threshold necessary for observing the FLASH effect. Subsequent experiments maintaining a fixed dose of 15 Gy while varying MDR from 40 to 750 Gy/s revealed that a significant reduction in histological scores emerged at the threshold MDR of 100 Gy/s (corresponding to 150 ms irradiation duration) (Figure 1C). Furthermore, the sparing efficacy exhibited progressive strengthening with increasing MDR, until reaching saturation above 200 Gy/s (Figure 1C). The impact of pulse structure was further investigated by maintaining a constant MDR of 150 Gy/s while varying pulse dose (0.5 or 3 Gy per pulse). At a total dose of 15 Gy, the FLASH effect was consistently observed for both high (3 Gy/pulse) and low (0.5 Gy/pulse) pulse dose (Figure 1D). In contrast, no FLASH effect was detected at a sub-threshold total dose of 9 Gy regardless of changing pulse structure (Figure 1E), reinforcing that surpassing a critical total dose threshold is essential for observing the FLASH effect.

To further evaluate the dose threshold and the influence of delivery modality, a hybrid irradiation paradigm combining FLASH-RT and CONV-RT was implemented (Figure 1F). Comparative analysis of acute intestinal toxicity was conducted across varying FLASH-RT/CONV-RT dose combinations under the same total dose. The MDR were 750 Gy/s for FLASH-RT and 0.1 Gy/s for CONV-RT, with a beam interval of ~85 seconds. For clarity, irradiation groups are denoted using prefixes 'F' for FLASH-RT and 'C' for CONV-RT, with suffix numbers indicating dose in Gy (e.g., F15 = 15 Gy FLASH-RT; F9C6 = 9 Gy FLASH-RT followed by 6 Gy CONV-RT).

Figure 1G presents results for a fixed total dose of 15 Gy with FLASH-RT dose varying from 3 Gy to 9 Gy. A significant FLASH sparing effect was observed when the FLASH-RT dose exceeded 6 Gy. When the FLASH-RT dose reached 9 Gy, the median acute histopathology score showed a marginal increase compared to the 6 Gy group, though this difference did not reach a statistical significance. However, the sparing effect was abolished when the FLASH-RT dose was reduced to 3 Gy. The F9C6 and F3C12 groups exhibited comparable tissue damage levels to the F15 and C15 groups, respectively. Notably, at a total dose of 10 Gy, no significant difference in tissue damage was observed compared to CONV-RT regardless of FLASH-RT/CONV-RT dose partitioning (Figure 1H). As shown in Figure 1G, altering the delivery sequence of FLASH-RT and CONV-RT did not yield significant differences. Quantitative analysis of Ki-67 immunofluorescence revealed significantly higher percentages of Ki-67-positive cells in the F15 and F9C6 groups compared to C15 and F3C12 groups (Figure 1I), demonstrating enhanced proliferative capacity preservation in FLASH-irradiated intestinal tissues. In addition to acute toxic injury in the small intestine, long-term biological endpoints including survival rate and body weight were also monitored in mice (Figure 1K). The F15 and F9C6 groups maintained a survival rate exceeding 50% at 20 days post-irradiation, whereas the median survival times for the C15 and F3C12 groups were only 7 and 5 days, respectively, demonstrating significant survival benefit conferred by the FLASH sparing effect. All irradiated groups exhibited a reduction in body weight, reaching the nadir at day 6 post-exposure followed by progressive recovery, with no significant differences in minimal body weights observed between groups.

While previous studies have generally believed that FLASH-RT necessitates high-dose delivery to activate sparing effects, the present study provides the first experimental evidence using a combinatorial irradiation strategy that the dose of FLASH irradiation can be greatly reduced while maintaining the sparing efficacy, provided that the total dose reaches a critical threshold. This finding opens new perspectives for mechanism exploration and clinical translation of FLASH-RT.

**Table 1.** Physical parameters of irradiation for all experimental groups.

| Group | Dose (Gy) | MDR (Gy/s) | Pulse frequency (Hz) | Pulse dose rate (Gy/s) | No. of pulses | Pulse dose (Gy) |
|---|---|---|---|---|---|---|
| Figure 1B | 6/9/12/15 | 900 | 300 | $1.00*10^6$ | 2/3/4/5 | 3 |
| Figure 1C | 15 | 40/100/200/750 | 13.33/33.33/66.67/250 | $1.00*10^6$ | 5 | 3 |
| Figure 1D | 15 | 150 | 300 | $2.00*10^5$ | 30 | 0.5 |
| Figure 1D | 15 | 150 | 50 | $1.00*10^6$ | 5 | 3 |
| Figure 1E | 9 | 150 | 300 | $1.25*10^5$ | 18 | 0.5 |
| Figure 1E | 9 | 150 | 50 | $7.05*10^5$ | 3 | 3 |

| Group | | Dose (Gy) | MDR (Gy/s) | Pulse dose rate (Gy/s) | Total exposure time(s) | Total MDR (Gy/s) |
|---|---|---|---|---|---|---|
| F15 | FLASH | 15 | 750 | $1*10^6$ | 0.02 | 750 |
| F9C6/C6F9 | FLASH | 9 | 750 | $1*10^6$ | 145.01 | 0.1 |
| F9C6/C6F9 | CONV | 6 | 0.1 | $4*10^3$ | | |
| F6C9/C9F6 | FLASH | 6 | 750 | $1*10^6$ | 175.01 | 0.09 |
| F6C9/C9F6 | CONV | 9 | 0.1 | $4*10^3$ | | |
| F3C12/C12F3 | FLASH | 3 | 750 | $1*10^6$ | 205 | 0.07 |
| F3C12/C12F3 | CONV | 12 | 0.1 | $4*10^3$ | | |
| C15/C10 | CONV | 15/10 | 0.1 | $4*10^3$ | 150/100 | 0.1 |
| C9F1 | FLASH | 1 | 750 | $1*10^6$ | 175 | 0.05 |
| C9F1 | CONV | 9 | 0.1 | $4*10^3$ | | |
| C6F4 | FLASH | 6 | 750 | $1*10^6$ | 125.01 | 0.08 |
|  | CONV | 4 | 0.1 | $4*10^3$ | | |

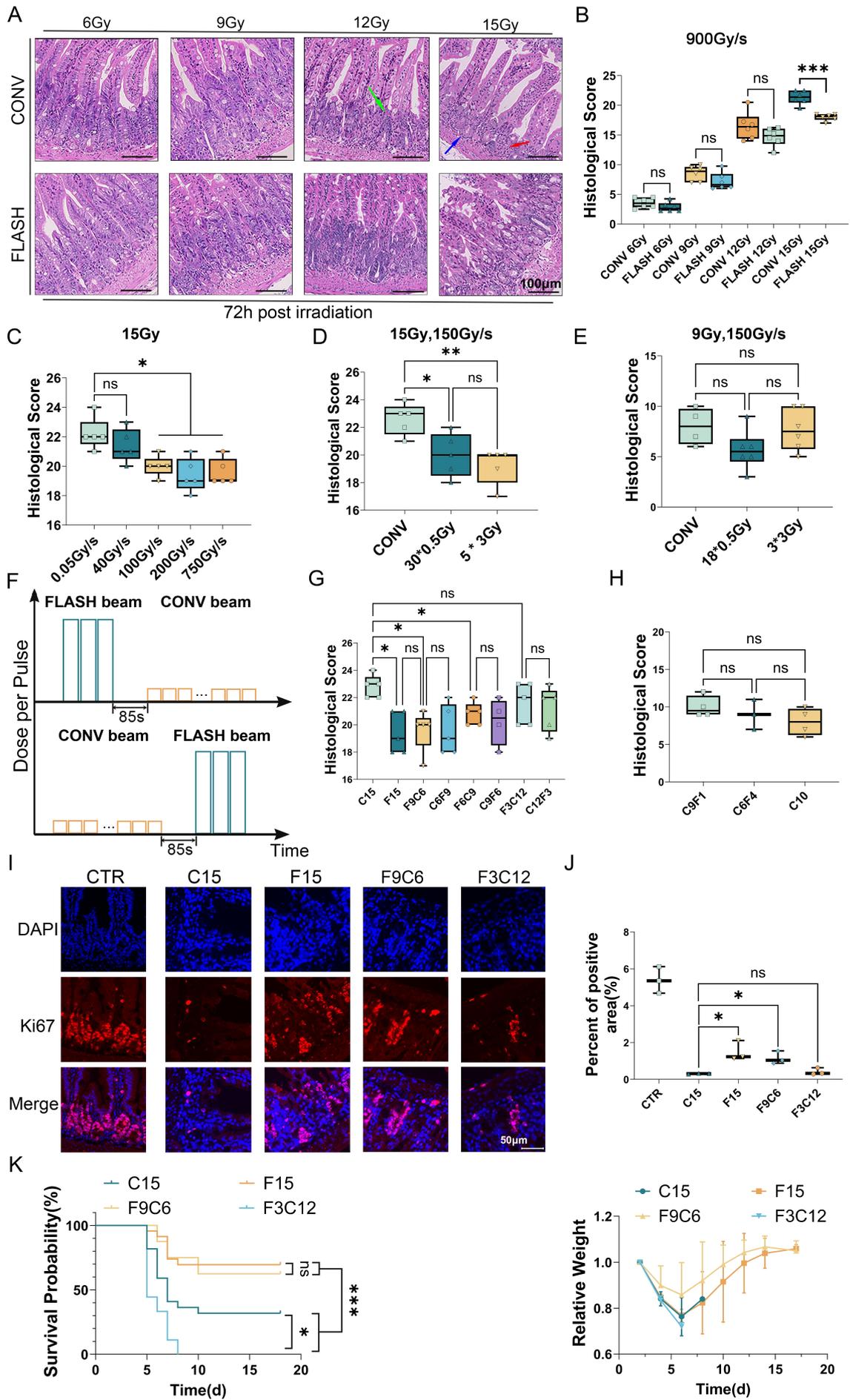

**Figure 1.** The physical factors of radiation required for the FLASH sparing effect were observed at the biological endpoint. (A) Representative images of hematoxylin and eosin (H&E) - stained sections of jejunum segments at 72h after different doses of whole abdomen irradiation (WAI). Different colors of arrows indicate different types of tissue injury mentioned previously, red: infiltration of inflammatory cells; blue: the architectural damage; green: a severe loss of numbers of crypts. Scale Bar= 100 μm. (B) Histological score of intestine tissues varied between FLASH-RT (900Gy/s) and CONV-RT across dose escalation (n=6 per group). (C) Changes of intestinal histological scores with different MDR after irradiation (n=5 per group). (D) Histological scores of intestinal tissues after irradiation at different pulse dose rates under the total dose of 15Gy (n=5 per group). (E) Histological scores of intestinal tissues after irradiation at different pulse dose rates under the total dose of 15Gy (n=4-6 per group). (F) Schematic of combined FLASH-RT and CONV-RT pulse sequences. (G) Intestinal tissue scores under combined CONV and FLASH beam irradiation at total doses of 15 Gy (n=4-5 per group). (H) Intestinal tissue scores under combined CONV and FLASH beam irradiation at total doses of 9 Gy (n=3-4 per group). (I) Representative immunofluorescence images of intestinal tissue sections co-stained with Ki-67 and DAPI. (J) Ki-67-positive percentage (n=3 per group, scale bar = 50 μm). (K) Kaplan-Meier survival curves and body weight change curves under different irradiation modalities (n=10 per group, log-rank (Mantel-Cox) test). p values are derived from *$p<0.05$, **$p<0.01$, ***$p<0.001$, One-way ANOVA test and Student's t test.

## 2.2. Tumor control efficacy: equivalent outcomes between CONV-RT and FLASH-RT

The FLASH effect refers to the mitigation of normal tissue toxicity while maintaining tumor control efficacy comparable to CONV-RT. To verify the tumor control efficacy of different irradiation modalities, tumor volume and mouse body weight were continuously monitored following radiation exposure. Tumors were irradiated with a total dose of 20 Gy using either FLASH-RT or CONV-RT alone. As shown in **Figure 2**A, both FLASH-RT and CONV-RT induced significant tumor growth arrest compared to the control group, demonstrating comparable tumor control efficacy. Besides, irradiation at a total dose of 15 Gy was administered to evaluate tumor response across different delivery schemes (F15, F9C6, F3C12 and C15). As illustrated in Figure 2B, all irradiated groups demonstrated comparable tumor control efficacy, inducing significant growth suppression or even regression compared to the control group. No significant changes in body weight were observed in irradiated mice compared to the control group (Figure 2C). Collectively, these results demonstrate that FLASH-RT achieves tumor control efficacy comparable to that of CONV-RT.

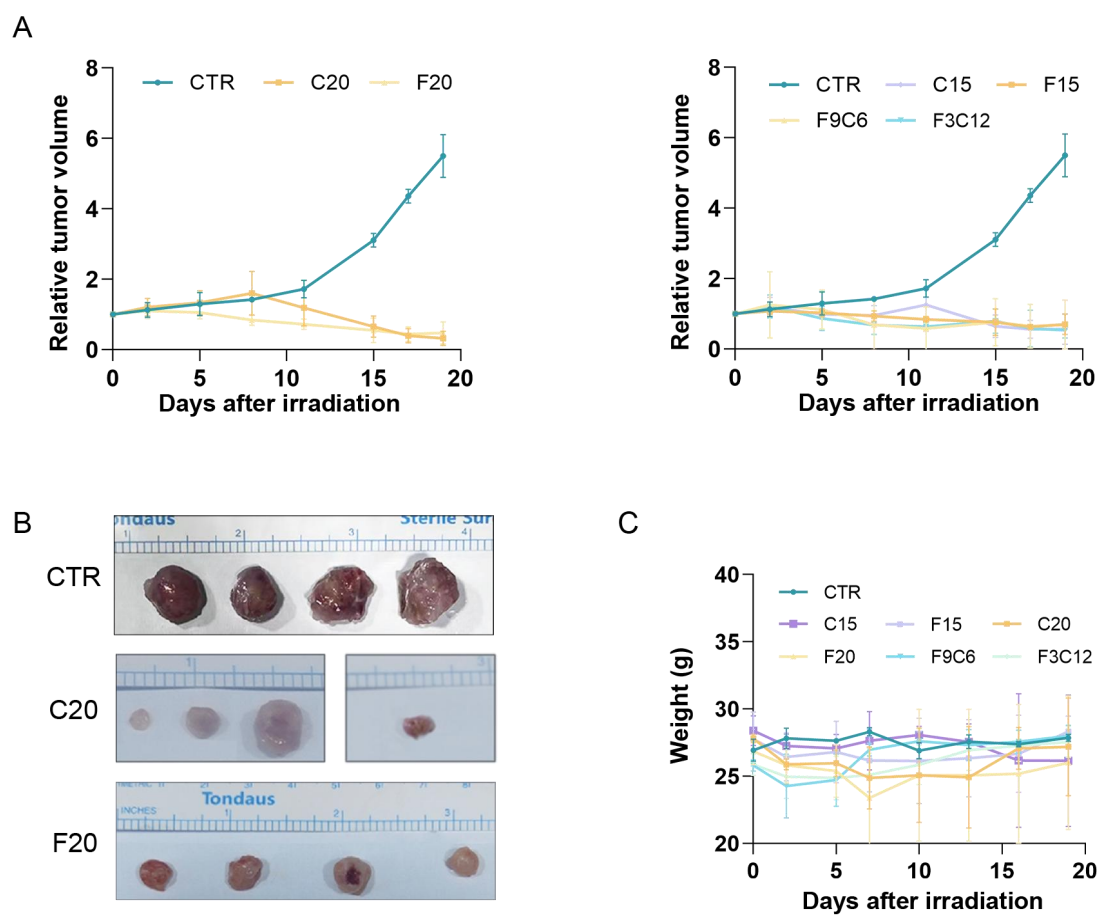

**Figure 2.** Comparison of tumor control efficacy between FLASH-RT and CONV-RT. (A) Tumor volume change curves under; Right panel: tumor volume change curves under different dose delivery schemes (F15, F9C6, F3C12) at 15 Gy total dose (n=4 per group). (B) Dissected tumors from the F20, C20, and CTR groups 18 days after irradiation (n = 4 per group). (C) Body weight change curves for the six experimental groups over the 18-day post-irradiation period (n=4 per group).

## 2.3. Critical Role of Radical in Triggering the FLASH sparing Effect

After elucidating the dose threshold characteristics of the FLASH effect, the mechanisms underlying the sparing effects were further investigated from both radiochemical reactions and biological responses. When ionizing radiation interacts with cellular components or water molecules, it generates free radicals, which are subsequently consumed either through radical recombination or via reactions with biological macromolecules such as lipids, proteins and DNA to form peroxides. To elucidate the underlying mechanisms of the FLASH sparing effect from a free radical perspective, a kinetic model was employed to simulate the chemical processes, specifically including the formation, recombination and scavenging of peroxyl radicals (ROO•). The influence of critical irradiation parameters (e.g., total dose, dose rate) on radical recombination was systematically evaluated.

The simulation results (**Figure 3**A-D) revealed a striking divergence between irradiation modalities: while the CONV group exhibited negligible radical recombination ratios (<0.001%) across all tested doses, the FLASH group demonstrated dose-dependent escalation of recombination efficiency. This suggests that a threshold dose must be reached to achieve a sufficient level of radical recombination, thereby eliciting an observable biological difference. As shown in Figure 3B, at a fixed dose of 15 Gy, the dose delivery duration profoundly influenced the radical recombination ratio: shorter irradiation times led to an increase in recombination ratio, which reached a saturation level when the time was reduced below 10 ms. Notably, the dose delivery time corresponding to 50% of the maximum radical recombination efficiency was approximately 300 ms.

Furthermore, the levels of ROOH following irradiation were also simulated. ROOH represents not only a product of free radical interactions with biomacromolecules but also a key mediator in oxidative stress and cellular damage, thereby serving as a well-established biomarker for radiation-induced injury. As shown in Figure 3C, the yield of ROOH increased with dose in both the CONV-RT and FLASH-RT groups. However, a progressive divergence emerged beyond 6 Gy, with FLASH-RT demonstrating significantly attenuated ROOH accumulation compared to CONV-RT at equivalent doses. The ROOH yields under various composite irradiation conditions at a fixed total dose of 15 Gy were further simulated. Using C15 group as the reference, the relative reduction in ROOH yield under different irradiation scenarios was calculated, as summarized in Figure 3D. The F15 group exhibited a reduction of 22.42%, while the F9C6 and F6C9 groups showed reductions of 8.79% and 3.95%, respectively. In contrast, the reduction in the F3C12 group was only 0.83%. In experimental groups where no FLASH effect was observed, the simulated reduction in ROOH yield was less than 3% compared to C15. Together, these findings demonstrate that radical recombination and peroxide levels play a crucial role in the FLASH sparing effect, providing a foundation for further mechanistic investigation.

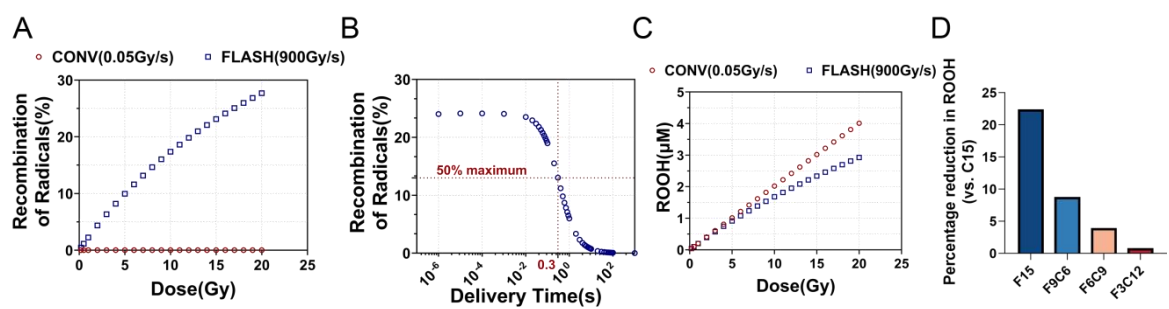

**Figure 3.** Simulation results of ROOH concentrations and radical recombination rates following CONV-RT and FLASH-RT. (A) Comparison of radical recombination proportions at varying doses between CONV-RT (0.05 Gy/s) and FLASH-RT (900 Gy/s). (B) Radical recombination proportion change curves versus delivery time at 15 Gy total dose. (C) Comparison of ROOH yields at varying doses between CONV-RT (0.05 Gy/s) and FLASH-RT (900 Gy/s). (D) Comparison of percentage decrease in ROOH yields relative to C15 for combined irradiation modes (F9C6, F6C9, F3C12) and F15.

Further investigation was conducted at the cellular and tissue levels to measure the levels of cellular ROS and Malondialdehyde (MDA), a product of lipid peroxidation, following irradiation. At a total dose of 15 Gy, the MDA production in the C15 group was significantly higher than that in the F15, F9C6, and F6C9 groups, while no significant difference was observed between the F3C12 and C15 groups (**Figure 4**A). Furthermore, two different human normal tissue cell lines (human mammary epithelial cell line MCF10A and small intestinal epithelial cell line FHS74Int) were employed for validation. At 1 hour post-irradiation with a total dose of 6 Gy, it was observed that intracellular ROS and MDA production induced by C6 were higher than those by F6 (Figure 4B and 4H). This indicates that at both tissue and cellular levels, the levels of free radicals and peroxides generated by FLASH-RT are lower than those by CONV-RT.

To investigate the role of free radicals in the FLASH effect, this study introduced antioxidant drugs to simulate the sparing effect of FLASH-RT and examined their alleviating effects on acute radiation-induced intestinal injury. At the tissue level, the antioxidant amifostine was employed, whose active metabolite (WR-1065) protects normal cells from radiation-induced DNA damage by scavenging free radical.[32,33] At the dose of 15 Gy, the differences in biological effects between the FLASH-RT group (MDR = 750 Gy/s) and the CONV-RT group (MDR = 0.05 Gy/s) were compared before and after amifostine intervention. As shown in Figure 4C, histological analysis after amifostine treatment, both irradiated groups exhibited significantly improved crypt structural integrity and reduced inflammatory cell infiltration. No significant differences were observed between the CONV combined with amifostine treatment group and the F15 group (Figure 4D). Further quantitative analysis of cell proliferation levels by Ki-67 immunofluorescence staining revealed that amifostine treatment significantly enhanced the proliferative capacity of the C15 group, with no significant difference in Ki-67 expression levels compared to the F15 group (Figure 4E and 4F). Notably, combined FLASH-RT and amifostine treatment resulted in a further increase in Ki-67 expression (Figure 4F), indicating a synergistic effect of the antioxidant and FLASH-RT in promoting cell proliferation.

Similarly, we conducted validation at the cellular level using N-Acetylcysteine (NAC) treatment. As a precursor of cysteine, NAC significantly increases intracellular glutathione (GSH) levels and exerts cytoprotective effects by directly scavenging free radicals and enhancing the endogenous antioxidant defense system.[34] This compound has been widely used in cellular experiments to antagonize oxidative stress injury, regulate apoptosis, and modulate inflammatory pathways.[34] Measurement of cell viability 24 hours post-irradiation using the Cell Counting Kit-8 (CCK-8) assay revealed a significantly higher survival rate in the F6 group than in the C6 group (Figure 4G). NAC pretreatment significantly improved the survival rate of the C6 group, with no statistical difference compared to the F6 group. NAC pretreatment did not significantly affect the survival rate of the F6 group. In fact, as illustrated in Figure 4H, a 12-hour NAC incubation prior to conventional dose rate irradiation significantly reduced MDA content compared to the C6 group, with no significant difference from the F6 group combined with or without NAC pretreatment. Collectively, these results at both tissue and cellular levels indicated that FLASH-RT effectively promoted cell survival and proliferation by reducing radiation-induced free radical and peroxidation level.

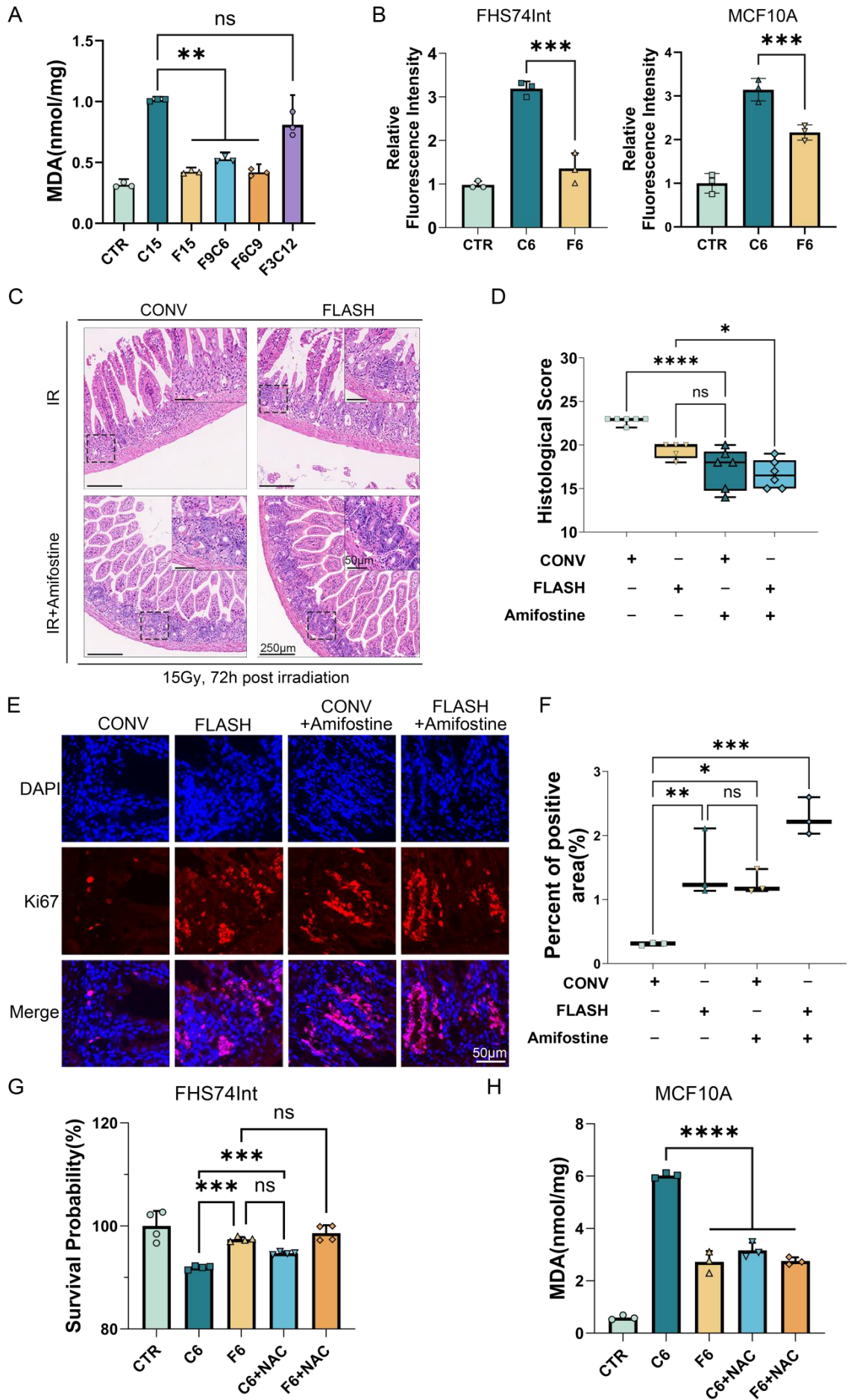

**Figure 4.** Comparative analysis of intestinal damage in mice subjected to amifostine administration prior to FLASH-RT or CONV-RT. (A) Intestinal tissue MDA levels across irradiation modalities (n=3 per group). (B) Relative ROS production in MCF10A and FHS74Int cells 1h post-CONV-RT or FLASH-RT (n=3 per group). (C) Representative images of H&E-stained sections of intestinal tissues 72 hours post-treatment, with or without amifostine, at irradiation dose rates of 750 Gy/s and 0.05 Gy/s, Scale Bar = 250μm (D) Histological scores of mice following irradiation, with or without the administration of amifostine (n=5-6 per group). (E) Representative immunohistochemical images of Ki-67 expression in intestinal tissues from the CONV-RT, CONV-RT + amifostine, FLASH-RT, and FLASH-RT + amifostine groups. Scale Bar = 50 μm. (F) Ki-67-positive percentage from the CONV-RT, CONV-RT + amifostine, FLASH-RT, and FLASH-RT + amifostine groups (n=3 per group). (G) CCK-8 cytotoxicity assay in FHS74Int cells 24h post-6Gy irradiation (n=4 per group). (H) MDA levels in MCF10A cells ± NAC pretreatment 1h after 6Gy irradiation (n=3 per group). p values are derived from *$p<0.05$, **$p<0.01$, ***$p<0.001$, one-way ANOVA test and Student's t test.

## 2.4. Comparative Analysis of Antioxidant Defense and Apoptosis in FLASH-RT versus CONV-RT

Previous studies indicated that the FLASH sparing effect is closely linked to cellular redox status. To investigate its molecular mechanisms, RNA-seq was conducted on intestinal tissues from mice irradiated with 15 Gy. Five experimental groups were included in the study: Untreated control group (CTR), Conventional dose-rate irradiation (C15), FLASH irradiation (F15), amifostine pretreatment + conventional dose-rate irradiation (C15+A), and Combined irradiation groups (F6C9 and F3C12). For irradiated groups, those exhibiting a sparing effect, namely F15, F6C9, and C15+A, were collectively designated as the Sparing Effect Group (SEG), while those without a sparing effect, C15 and F3C12, were classified as the No Sparing Effect Group (NSEG). Differentially expressed genes (DEGs) were analyzed using the criteria $|\log_2 FoldChange| \geq 1$ and $p < 0.05$. As shown in **Figure 5**A and Figure S2, supporting information, compared to the CTR group, the irradiated groups exhibited a large number of upregulated and downregulated DEGs. In contrast to the NSEG group, the SEG group manifested upregulation of genes associated with antioxidant defense and damage repair, while genes related to apoptosis, pyroptosis, and inflammatory signaling were downregulated. KEGG enrichment analysis highlighted that radiation exposure in C15 and F15 groups led to the activation of NF-κB and cytokine-cytokine receptor interaction pathways, indicative of heightened tissue inflammation. Additionally, upregulation of PI3K-AKT and MAPK pathways implied activation of compensatory mechanisms against oxidative damage (Figure 5B). Notably, the SEG group, compared to NSEG, demonstrated enhanced glutathione metabolism and suppression of MAPK and inflammatory pathways (Figure 5B), suggesting these pathways are central to mediating the FLASH effect.

The glutathione metabolism pathway serves as a cornerstone of the cellular antioxidant defense and detoxification systems, incorporating a series of GST family members and other antioxidant genes. As shown in the heatmap (Figure 5C), these antioxidant genes were significantly upregulated in the SEG group. To further validate these findings, real-time polymerase chain reaction (RT-qPCR) was performed to assess the mRNA expression levels of key antioxidant genes, including *Gsta1* and *Hmox1*, in tissues at 24 hours post-irradiation. As illustrated in Figure 5D, the mRNA expression levels of both *Gsta1* and *Hmox1* were significantly lower in the NSEG group compared to the SEG group. Furthermore, at 8 h, 24 h, 72 h, and 6 days post-irradiation, the mRNA expression of *Gsta1* and *Hmox1* in the F15 group remained consistently higher than in the C15 group (Figure 5E). These findings suggest that the FLASH effect is closely associated with the enhanced activation of antioxidant defense mechanisms in response to radiation-induced intestinal toxicity.

The glutathione system, catalase (CAT), and superoxide dismutase (SOD) constitute key redox regulatory systems within cells, essential for maintaining intracellular redox homeostasis. We further measured the production of reduced GSH and oxidized glutathione (GSSG) in the FHS74Int cell line at 8 hours post-6 Gy irradiation, using the GSH/GSSG ratio as an indicator of intracellular redox status. Figure 5F demonstrates that the GSH/GSSG ratio was significantly higher in the F6 group compared with the C6 group. Concurrently, CAT enzyme activity was also markedly elevated in the F6 group, compared with the C6 group (Figure 5G). These results collectively indicated that FLASH-RT conferred a stronger antioxidant defense capacity. Additionally, NAC pretreatment elevated both the GSH/GSSG ratio and CAT activity in the C6 group to levels comparable to those in the F6 group (Figures 5F and 5G). These findings further indicated that the sparing effect of FLASH-RT was attributable to its ability to reduce the generation of free radicals and peroxides, thereby efficiently restoring the cellular antioxidant defense capacity.

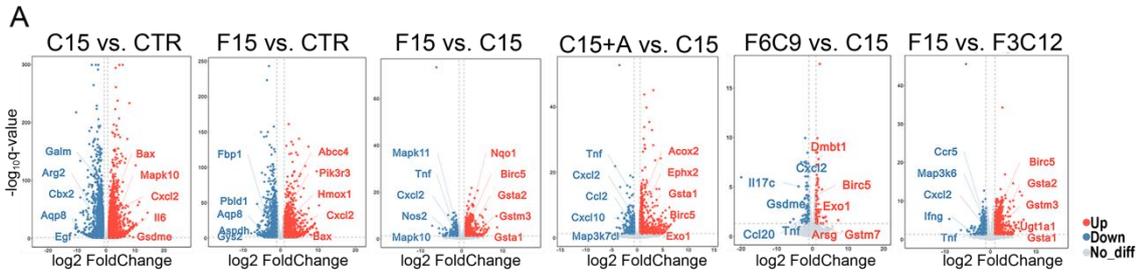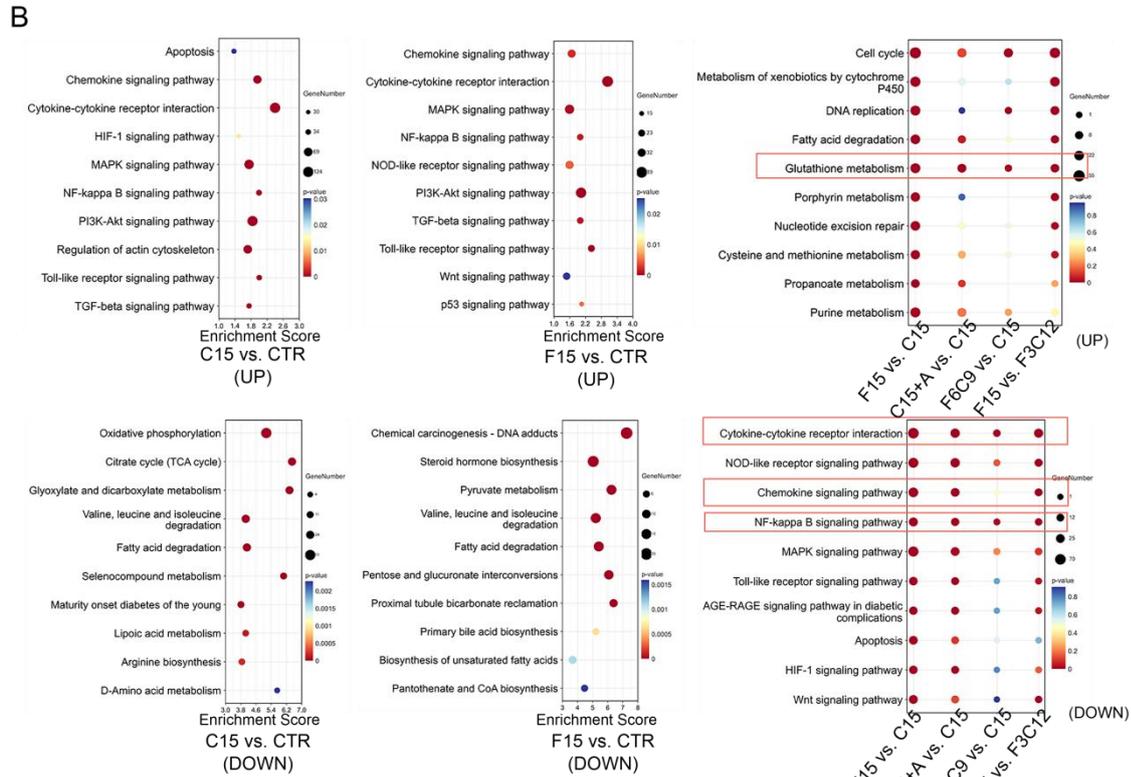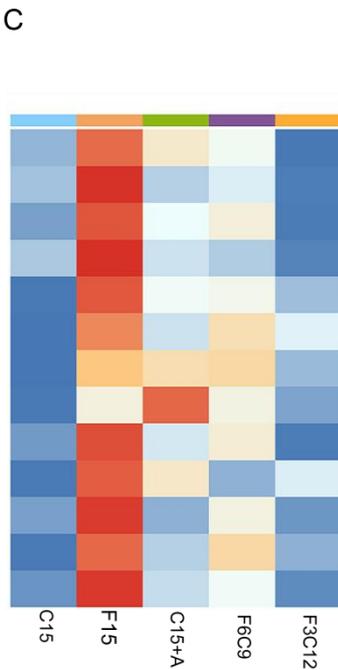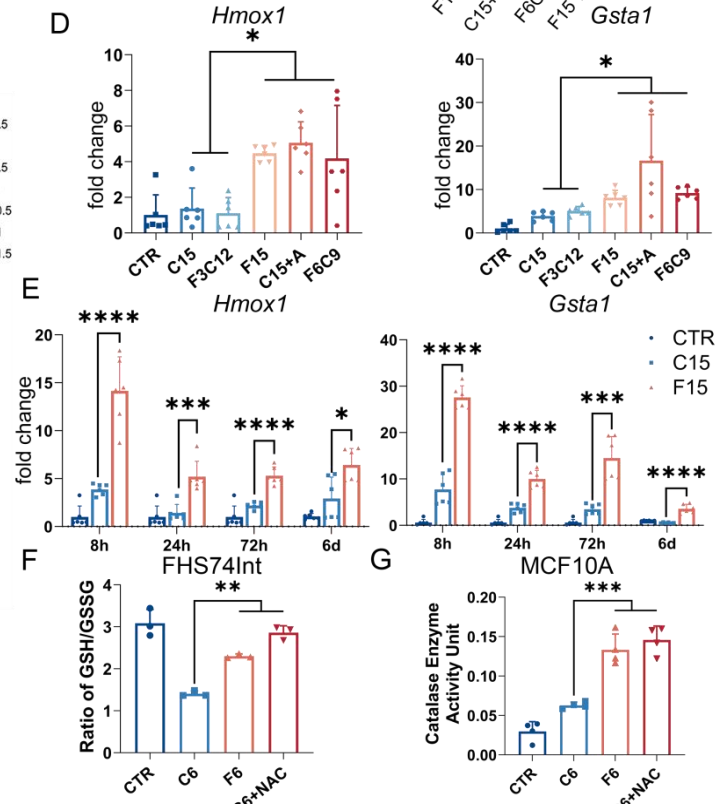

**Figure 5.** RNA-seq analysis reveals that FLASH irradiation upregulates glutathione metabolism and enhances antioxidant enzyme activity compared to conventional radiotherapy. (A) Volcano plots of differentially expressed genes between groups (from left to right: C15/CTR, F15/CTR, F15/C15, C15+A/C15, F6C9/C15, F15/F3C12; groups after '/' serve as controls).(B) KEGG enrichment analysis results of the aforementioned differentially expressed genes. (C) Heatmaps of antioxidant gene expression levels across different irradiation modalities. (D) Validation of *Hmox1* and *Gsta1* mRNA levels in murine intestinal tissue 24h post-15 Gy irradiation under various dose delivery schemes via RT-qPCR (n=6 per group). (E) Temporal changes in *Hmox1* and *Gsta1* mRNA levels in murine intestinal tissue following 15 Gy CONV-RT or FLASH-RT at specified time points (8h, 24h, 72h, 6d) (n=6 per group). (F) Measurement of GSH and GSSG in FHS74Int cells 8h post-6Gy CONV-RT or FLASH-RT, with ratio comparison (n=3 per group). (G) CAT activity in MCF10A cells 8h post-6Gy CONV-RT or FLASH-RT (n=3 per group). p values are derived from *p<0.05, **p<0.01, ***p<0.001, ****p<0.0001, one-way ANOVA test and Student's t test..

To elucidate the mechanisms underlying the enhanced antioxidant capacity and attenuated radiation damage in the SEG group, this study pursued two lines of investigation. First, we focused on the NRF2 pathway, a central regulator of antioxidant defense, which directly modulates the glutathione metabolism pathway as previously described. Second, we examined the differences in cellular apoptosis between FLASH-RT and CONV-RT.

Previous studies have demonstrated that ROS activate the transcription factor NRF2, thereby initiating the cellular antioxidant defense program.[35] This study investigated the expression levels of NRF2, p-NRF2, and its downstream protein HO-1. In intestinal tissues, at 24 h post-irradiation, p-NRF2 expression was higher in the F15 group compared to the C15 group (**Figure 6**A). However, no significant difference in HO-1 protein expression was observed between the F15 and C15 groups at this time point (Figure 6A). Further analysis of HO-1 expression at multiple time points revealed that HO-1 levels were significantly elevated in the F15 group compared to the C15 group at 2 h and 4 h post-irradiation (Figure 6B), indicating time-dependent fluctuations in protein expression following irradiation.

Additionally, in FHS74Int cells at 12 h post-irradiation, both p-NRF2 and HO-1 were significantly upregulated in the F6 group relative to the C6 group (Figure 6C). To confirm the critical role of the NRF2 antioxidant pathway in cell survival, NRF2 was knocked down in FHS74Int cells, with knockdown efficiency validated by Western blot (Figure S3, Supporting Information). Comparison of cell survival post-irradiation between NRF2-knockdown and wild type cells showed that NRF2 depletion markedly reduced cell viability and abolished the survival difference between the C6 and F6 groups (Figure 6D). These results suggest that FLASH-RT mitigates tissue damage by significantly activating the NRF2 antioxidant defense pathway following irradiation.

Furthermore, the sequencing results revealed significant downregulation of the MAPK signaling pathway in the SEG group (Figure 5B). Within the MAPK pathway, the classical ERK pathway is one of the key regulators of apoptosis. As shown in Figure 6E, the phosphorylation level of ERK in the F15 and C15+A groups was significantly lower than that in the C15 group. Consistently, the level of cleaved-caspase3 was also markedly reduced in these groups compared to the C15 group (Figure 6F). These findings indicated that FLASH-RT exerted an anti-apoptotic effect similar to that of antioxidant agents. In summary, FLASH-RT enhanced the antioxidant defense capacity and reduced apoptosis in intestinal tissues by modulating the NRF2 antioxidant pathway and the MAPK pathway, thereby mitigating radiation-induced tissue damage.

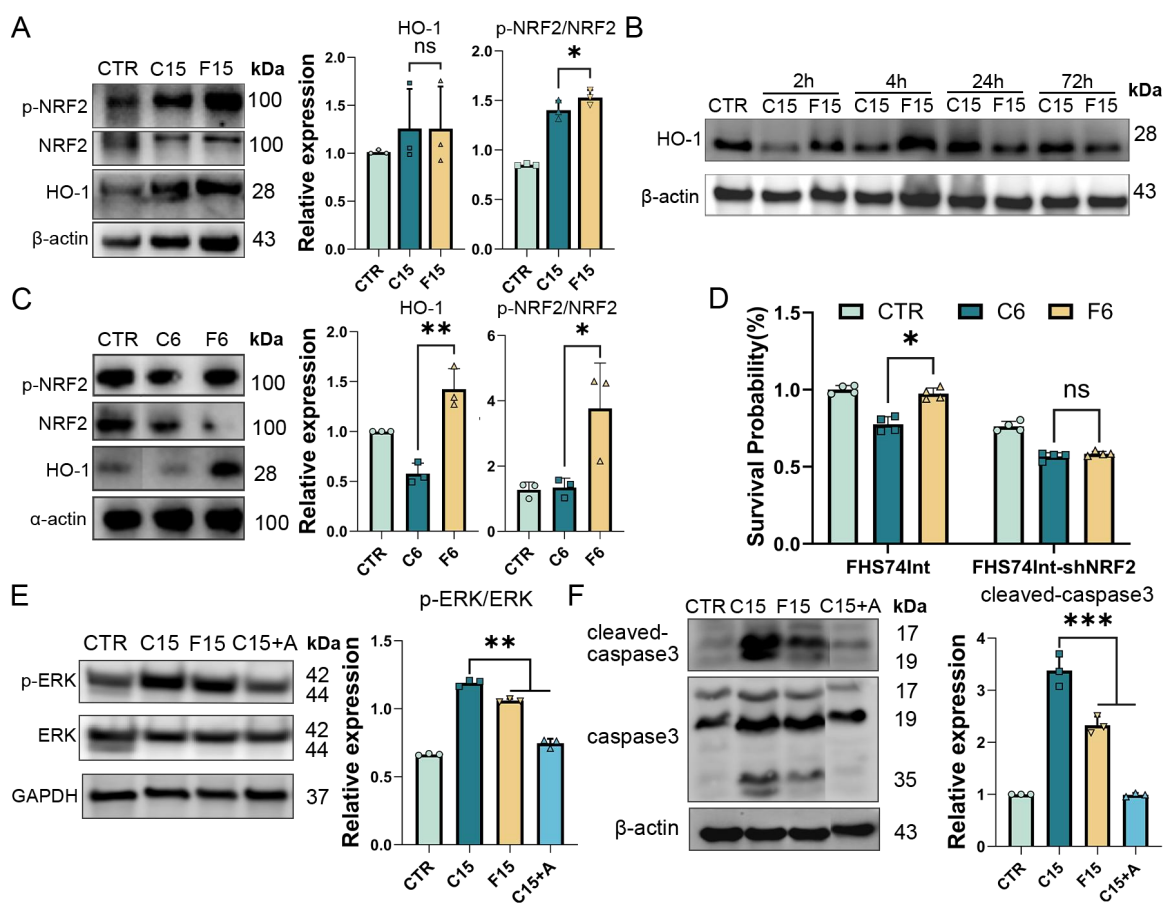

**Figure 6.** Comparative analysis of NRF2-mediated antioxidant defense and apoptosis in FLASH-RT versus CONV-RT. (A) Protein levels of NRF2, p-NRF2, and HO-1 in small intestinal tissue 24h post-C15 versus F15 treatment with band densitometry analysis (n=3 per

group). (B) The expression levels of HO-1 in the C15 and F15 groups at 2h, 4h, 24h, and 72h after irradiation (n=3 per group). (C) p-NRF2 and HO-1 levels in FHS74Int cells 12h after C6 or F6 treatment with band densitometry analysis (n=3 per group). (D) Cell viability comparison in CTR and NRF2-knockdown FHS74Int cells subjected to C6 or F6 irradiation (n=4 per group). (E) ERK phosphorylation in intestinal tissue 24h post-treatment across C15, F15, and C15+A groups (n=3 per group). (F) Expression levels of cleaved-caspase3 in intestinal tissue 24h post-treatment across C15, F15, and C15+A groups (n=3 per group). p values are derived from *p<0.05, **p<0.01, ***p<0.001, Student's t test.

## 3. Discussion

To date, no unified consensus has been reached regarding the definitive physical parameters required to observe the FLASH effect at biological endpoints. Existing literature reports inconsistent findings on the influence of various physical factors (such as dose, MDR, and pulse dose rate) on the manifestation of the FLASH effect.[7,8,14,36,37] Our statistical analysis of published literature studies indicates that both dose and MDR must exceed certain thresholds to elicit the FLASH effect.[13] Given the pivotal role of free radicals in mediating biological damage induced by low-LET radiation, this study investigates the sparing mechanisms of the FLASH effect from the perspective of free radical dynamics.[20] Although prior studies have proposed hypotheses such as the radical recombination hypothesis and radical recombination-antioxidant hypothesis, systematic experimental validation in complex biological entities, such as cells or tissues, remains lacking. Therefore, this study focuses on the role of free radicals, employing a combined simulation and experimental approach to elucidate the key factors and underlying mechanisms that induce the FLASH effect. This integrated framework establishes the foundation for our subsequent investigations into the modulation of redox homeostasis and its impact on optimizing FLASH delivery parameters.

Simulations results indicate that under CONV-RT (MDR = 0.05 Gy/s), minimal radical recombination occurs, with ROOH increasing linearly with dose (Figure 3A). In contrast, for FLASH-RT (MDR=900Gy/s), radical recombination probability rises significantly with escalating doses, leading to a progressive divergence in ROOH levels compared to CONV-RT. These simulations suggest that a threshold dose is required to trigger sufficient radical recombination, thereby generating observable biological differences. This inference is supported by our experimental findings (Figure 1B), which show no FLASH effect was observed at 6, 9, or 12 Gy under a MDR of 900Gy/s, until a clear effect emerges at 15 Gy.

Notably, this 15 Gy threshold closely approximates the lethal dose (LD50) (14.7 Gy) for acute intestinal toxicity in mice.[6] Similarly, previous studies have reported that the dose threshold for FLASH sparing effect in skin tissue is 25 Gy, which aligns with the LD50 range for skin (25–28 Gy). These observations imply that the FLASH effect dose threshold may be directly correlated with tissue radiosensitivity.[10,13,38] To comprehensively evaluate radiation damage, we employed histopathological scoring as an integrated assessment metric. The results revealed that 15 Gy FLASH-RT yielded a histological score of 20, equivalent to the damage level induced by 12 Gy CONV-RT (Figure 1B). Simulations further demonstrated that these two irradiation conditions produced comparable peroxide levels (Figure 3C). This finding provides robust evidence that radical recombination is the key mechanism underlying the radioprotective FLASH effect.

For a given dose, the radical recombination efficiency depends on the temporal profile of dose delivery. As shown in Figure 3B, simulations of radical recombination kinetics at 15 Gy reveal that recombination remains negligible under prolonged delivery times (i.e., conventional dose rates), whereas it approaches saturation when the delivery window is compressed below 1 ms. The time window corresponding to 50% maximal recombination efficiency is approximately 330 ms. This prediction aligns closely with our experimental findings (Figure 1C), where no FLASH effect was observed at a MDR of 40 Gy/s (delivery time: 375 ms), but became evident at 100 Gy/s (150 ms). This consistency between simulation and experiment further confirms that, at a fixed dose, the MDR is a critical determinant for inducing the FLASH effect. This conclusion is corroborated by independent studies: Sunnerberg et al. identified MDR as the key parameter governing radiation-induced oxygen consumption and hydrogen peroxide generation in albumin solutions.[39] Similarly, Grilj et al. reported significant dose-rate-dependent trends in radiation-induced lipid peroxidation in liposomal systems.[14]

Furthermore, under fixed total dose and MDR conditions, variations in pulse dose and pulse number yielded no significant differences in sparing effect (Figure 1D). Notably, FLASH sparing effect was still observed even with pulse doses as low as 0.5 Gy/pulse, provided the repetition frequency was increased to achieve high MDR (e.g., 150 Gy/s as tested in this study). This finding aligns with previous reports.[14] In contrast, a few studies have reported sparing effects at high pulse doses (≥4.7 Gy/pulse) despite near-conventional MDR.[7] We hypothesized that at high single-pulse doses, the instantaneous dose rate (reaching MGy/s levels) may induce sufficient radical recombination within a single pulse to trigger FLASH sparing effect.

To test this hypothesis, we innovatively designed a combined FLASH-RT and CONV-RT irradiation protocol (Figure 1F), adjusting the dose ratio between FLASH and CONV components to evaluate acute intestinal toxicity. As shown in Figure 1H, at a total dose of 15 Gy, FLASH effect became evident when the FLASH component exceeded 6 Gy. However, at 10 Gy total dose, no sparing effect was observed regardless of the FLASH/CONV dose ratio (Figure 1H). Moreover, the sequence of FLASH-RT and CONV-RT delivery did not significantly affect outcomes (Figure 1G). Notably, the overall MDR in these combined-beam experiments ranged only from 0.07–0.1 Gy/s (Table 1), suggesting that the conventional definition of 'MDR' (total dose divided by total time) may be inadequate for predicting FLASH effects.

Indeed, the F6C9 group exhibited significantly lower MDA levels compared to the C15 group (Figure 4A), further underscoring the pivotal role of free radicals in mediating the FLASH effect. This was further corroborated by antioxidant intervention studies. The addition of antioxidants competes with radical recombination for free radical binding,[40] resulting in reduced MDA production (Figure 4H). Pretreatment with antioxidants markedly attenuated acute intestinal injury in the CONV group (Figure 4D), while significantly enhancing proliferative capacity and cell survival to levels comparable to the FLASH-RT group (Figures 4F and 4G). In the FLASH-RT group, antioxidant administration did not substantially alter injury severity but further increased the Ki-67 positive ratio, indicating enhanced proliferative potential. Moreover, tumor tissues exhibit higher antioxidant concentrations and antioxidant enzyme activities than normal tissues,[41] which may explain the equivalent tumor control efficacy between FLASH and CONV-RT despite differential radioprotection in normal tissues. These findings suggest that the therapeutic window of FLASH-RT may be further optimized by modulating tissue redox status. Additionally, the differential response of normal versus tumor tissues to antioxidant modulation warrants investigation in clinical FLASH-RT applications.

Excessive or persistent oxidative stress induces cellular damage. NRF2, a master antioxidant transcription factor, protects cells from oxidative damage through rapid nuclear accumulation of p-NRF2, which enhances ARE-driven expression of genes encoding antioxidant defense proteins and detoxification enzymes.[42,43] Our findings demonstrate that the F15 group exhibited a significantly higher p-NRF2/NRF2 ratio compared to the C15 group, accompanied by elevated expression of its downstream protein HO-1 (encoded by *Hmox1*) (Figure 6A, 6C). Concurrently, transcriptional levels of *Hmox1* and *Gsta1* were markedly upregulated in the F15 group (Figure 5D, E). HO-1 not only modulates redox activity but also regulates iron homeostasis.[44,45] As a member of the glutathione S-transferase (GST) family, *Gsta1* upregulation indicates enhanced detoxification capacity (Figure 5D, E).[46] These results collectively suggest that the F15 group restores cellular redox homeostasis more effectively through NRF2-mediated enhancement of antioxidant and detoxification responses, which is further corroborated by the higher GSH/GSSG ratio observed in the F15 group compared to the C15 group (Figure 5F).

Ionizing radiation can also activate redox-sensitive signaling kinases, including ERK and AKT,[47,48] which modulate protein activity through phosphorylation. Our results revealed that the F15 group exhibited significant enrichment of the MAPK signaling pathway (Figure 5B) accompanied by markedly reduced ERK phosphorylation compared to the C15 group (Figure 6E). Consistent with reports that downregulation of p-ERK suppresses cell death and confers increased resistance to irradiation,[27] and the F15 group displayed lower levels of cleaved-caspase 3 than the C15 group (Figure 6F). These findings suggest that FLASH-RT may attenuate cell death by inhibiting ERK phosphorylation. Notably, the C15 group combined with antioxidant treatment showed outcomes comparable to the F15 group, further validating that reduced ROS levels contribute to ERK suppression and subsequent cytoprotection (Figure 6E).

Notably, NRF2 exhibits crosstalk with key signaling pathways such as MAPK and NF-κB.[49] Previous studies have demonstrated that p38 MAPK negatively regulates NRF2 activation. Specifically, p38 MAPK activation induces NRF2 phosphorylation and promotes its binding to KEAP1, potentially reducing NRF2 nuclear translocation.[50] Conversely, the antioxidant sulforaphane inhibits Trichloroethene (TCE)-induced phosphorylation of p38, which may disrupt the NRF2/KEAP1 complex, facilitating NRF2 nuclear translocation. This process induces cytoprotective genes to counteract TCE-mediated immune responses.[51] Furthermore, the interaction between NRF2 and NF-κB is identified as a critical mechanism underlying TCE-triggered inflammation and autoimmunity.[51] Inhibition of MAPK p38 also leads to reduced transcriptional activity of NF-κB.[52] Collectively, these findings suggest that the enhanced NRF2 activity observed in this study may contribute to alleviating inflammatory responses, representing a promising direction for future investigation.

## 4. Conclusion

In conclusion, this study elucidates the critical roles of dose, MDR, and pulse dose in triggering the FLASH sparing effect and provides an integrated understanding of their interrelationships. Through an innovative approach combining pulsed electron FLASH-RT and CONV-RT in a murine model of acute intestinal toxicity, we demonstrate for the first time that FLASH-RT can induce a robust sparing effect even at a reduced dose of 6 Gy, provided the total dose reaches a necessary threshold. Our findings reveal that FLASH-RT significantly enhances peroxyl radical recombination, reduces ROS levels, and lowers MDA content compared to CONV-RT, highlighting the pivotal role of free radical dynamics in the FLASH sparing effect. This conclusion is further supported by the observation that antioxidant administration effectively abolishes the differential outcomes between FLASH-RT and control groups at both cellular and tissue levels. Transcriptomic analysis further shows that both FLASH-RT and CONV-RT with antioxidant supplementation markedly upregulate glutathione metabolism pathways while suppressing inflammatory responses. Mechanistically, FLASH-RT confers tissue sparing by activating the NRF2 antioxidant pathway and inhibiting ERK signaling, thereby enhancing cellular redox defense, attenuating apoptosis, and ultimately mitigating ROS-mediated tissue damage. These findings underscore the feasibility of optimizing the therapeutic window of FLASH-RT through targeted modulation of tissue redox status. Furthermore, they establish a foundation for evaluating free radical-targeted strategies to enhance the efficacy of FLASH radiotherapy, with significant implications for its clinical translation.

## 5. Experimental Methods

*Mice Handling*: All mice were procured from the Laboratory Animal Resources Center at Tsinghua University (THU-LARC, accreditation number was 23-QR2) for experimental purposes. The mice were maintained on a diet of standard chow and had unrestricted access to acidified water. Housing conditions included cages with a regulated 12-hour light/dark cycle. All procedures involving the mice received approval from the Institutional Animal Care and Use Committee (IACUC).

To evaluate the FLASH sparing effect on normal tissues, female C57BL/6J mice (Strain code: N000013, GemPharmatech Co., Ltd., Nanjing, China), aged six to eight weeks old, were used in the experiments. Separately, a tumor-bearing mouse model was established using male BALB/c mice (6-8 weeks old, Strain code: N000020, GemPharmatech Co., Ltd., Nanjing, China) to assess tumor growth inhibition of CONV-RT and FLASH-RT.

*Irradiation Setup*: Irradiation was conducted utilizing an electron linear accelerator, which provides a vertically irradiated electron beam with an energy of 6 MeV. WAI was administered to the C57BL/6J mice by positioning a collimator at the accelerator's exit window, thereby achieving a field size of 45 × 45 mm² (refer to **Figure 7**A and 7B). The radiation dose for each mouse was quantified using calibrated EBT3 films (gafchromic^TM), with doses ranging from 6 to 15 Gy. For tumor irradiation, mice were equipped on the platform with the radiation field collimated to 10 × 10 mm², tumors were placed at the center of the radiation field, with normal tissues kept from exposure area during CONV-RT or FLASH-RT. During irradiation, mice were positioned on a board with a film placed both above and below the irradiation site. The films were scanned 24 hours post-irradiation, and the validation of homogeneous dose coverage is illustrated in Figure 7C and 7D. Figure 7E and 7F demonstrated that within the irradiation field, the doses administered in the vertical, longitudinal, and lateral directions achieved the planned levels of 8 Gy for CONV-RT and 16 Gy for FLASH-RT. The dose distribution was observed to be uniform. All conventional irradiation procedures were conducted at a dose rate of 0.05 Gy/s. The specific beam parameters for FLASH-RT are detailed in Table 1.

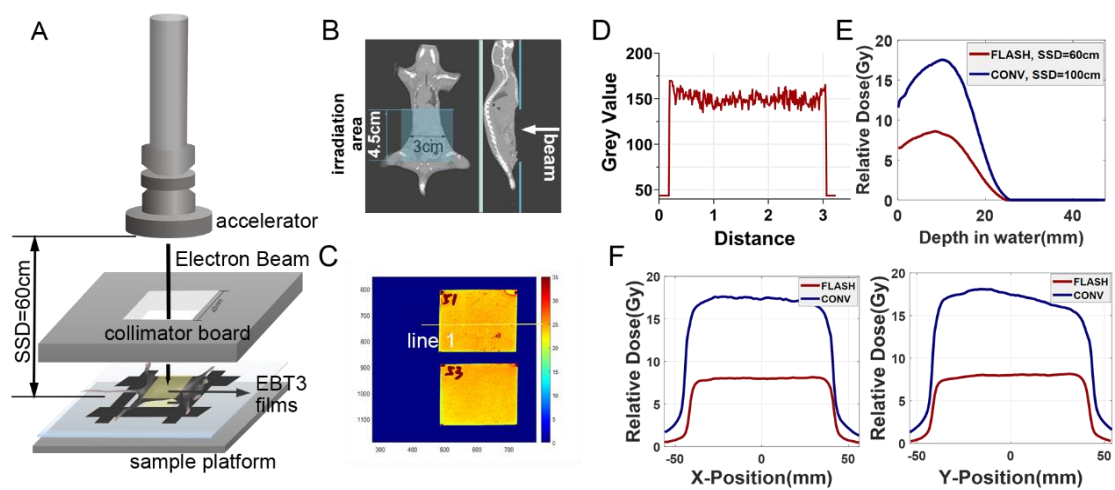

**Figure 7.** The experimental platform employed for WAI. (A) A schematic representation of the experimental setup for FLASH-RT. (B) Top and side computed tomography images of mice illustrating the irradiation status. (C) The spatial dose distribution was measured using an EBT3 film. (D) A straight line (line 1) was drawn on the film in Figure 7C to assess grey values at different points. (E) The vertical dose along the central axis of the irradiation plane. (F) The relative dose variations in the longitudinal and lateral directions within the plane. The planned doses were 8 Gy for CONV-RT and 16 Gy for FLASH-RT.

*Drug Treatment*: Amifostine (MCE,#HY-B0639) was prepared in sterile saline at 40 mg/mL and given intraperitoneally at 200 mg/kg. A prior study determined the median LD50 of amifostine to be 833±20 mg/kg, a value significantly exceeding the dosage employed in our experiment.[32] Consequently, the intrinsic toxicity of the drug can be considered minimal. Amifostine was administered 15-30 minutes prior to irradiation. For the cell experiments,

NAC (MCE,#HY-B0215) was dissolved in ultrapure water to prepare a 500 mM stock solution, which was then filtered through a 0.22μm filter. Cells were incubated with 10 mM of NAC for 12 hours prior to irradiation to inhibit radiation-induced ROS production.

*Survival Analysis*: Mice were monitored for 30 days post-irradiation, and then conducted euthanasia. Animals were checked daily for acute toxicity—including weight loss, hunching, diarrhea, or lethargy—and euthanized if their condition met predefined humane endpoints: sustained weight loss >15%, severely reduced activity, or compounded morbidity (e.g., >10% weight loss with diarrhea and lethargy). Euthanasia was performed by carbon dioxide asphyxiation followed by cervical dislocation. Survival data were analyzed by Kaplan–Meier curves and the log-rank test.

*Tissue Processing and Histological Analysis*: A five-centimeter segment of the jejunum was excised, rinsed in phosphate-buffered saline (PBS), and subsequently fixed overnight in 4% paraformaldehyde (PFA). The following day, the opened intestine was rolled into a Swiss roll configuration, paraffin-embedded, sectioned and H&E-stained. For histological analysis, a semi-quantitative histological score was determined based on a modified Geobes score and Chiu's score,[29,31] evaluating six parameters, including submucosal inflammation, crypt structure, crypt regeneration, inflammatory cell infiltration in the lamina propria, thickening of the muscularis mucosae and epithelial damage. Severity is graded on a scale from 0 (least severe) to 4 (most severe), as detailed in Table S1.

*Determination of Cell Viability and Proliferation*: For the Ki67 Immunofluorescence (IF) assay, paraffin sections were deparaffinized and subjected to antigen retrieval. After blocking with 5% normal goat serum, sections were incubated overnight at 4 °C with anti-Ki67 antibody (Servicebio,#GB111499). Following PBS washes, the sections were incubated with a fluorescently conjugated secondary antibody. Nuclei were counterstained with DAPI, and slides were mounted with antifade mounting medium for fluorescence microscopy analysis. For the CCK-8 assay, cells were seeded in 96-well plates, irradiated at 80% confluence, and cultured for 24 h. CCK-8 solution was added, followed by 2 h incubation at 37°C. Absorbance was measured at 450 nm, and viability was normalized to controls.

*RNA-Seq*: Mice intestinal tissues were collected three days after irradiation, then ground into powder in liquid nitrogen. Total RNA was extracted using an Animal RNA Isolation Kit with Spin Column (Beyotime, #R0026), and its concentration was measured using a NanoDrop spectrophotometer. mRNA was enriched from total RNA using poly(T) magnetic beads, fragmented, and used for first-strand and second-strand cDNA synthesis. The resulting cDNA underwent end repair, A-tailing, and adapter ligation. Finally, adapter-ligated cDNA fragments were amplified and purified prior to sequencing. Transcriptome sequencing and analysis were performed by OE Biotech Ltd (Shanghai, China).

*RT-qPCR*: Total RNA was extracted from intestinal tissues using the RNA Isolation Kit illustrated above. Using a NanoDrop One Spectrophotometer to assess RNA concentration. cDNA was then synthesized using TRUEscript RT MasterMix (OneStep gDNA Removal) to obtain the cDNA template. qPCR was performed using PowerUp SYBR Green Master Mix (Thermo Fisher Scientific, #A25742) on a QuantStudio™ 3 Real-Time PCR System. A 20 μsL qPCR reaction mixture was prepared with 10 μL SYBR Green Master Mix, 2 μL cDNA, and 200 nM of each primer. The protocol included UDG pre-incubation at 50 °C for 2 min, initial denaturation at 95 °C for 2 min, followed by 40 cycles of 95 °C for 15 s and 60 °C for 1 min. Gene expression was quantified using the comparative threshold cycle (ΔΔCt) method normalized to β-actin. Primer sequences are provided in Supplementary Table S2.

*Western Blot*: Total proteins were extracted from tissues and cultured cells using ice-cold RIPA lysis buffer containing Protease/Phosphatase Inhibitors (CST,#5872). Tissue samples were homogenized with a tissue homogenizer (-10°C, 5 min). Cells were lysed in RIPA buffer and sonicated. After incubation on ice and centrifugation, protein concentrations were determined by Bicinchoninic Acid (BCA) assay to ensure equal loading. Proteins were separated by 10% SDS-PAGE and transferred to PVDF membranes. The membranes were blocked with 5% non-fat milk in TBST and incubated overnight at 4 °C with primary antibodies (1:1000). Following TBST washes, membranes were incubated with HRP-conjugated secondary antibodies (1:5000) for 1 h at room temperature. Signals were detected using an ECL kit and visualized on a chemiluminescence imaging system. Band intensities were quantified with ImageJ.

The following primary antibodies were used: anti-NRF2 (Abcam,#ab62352), anti-p-NRF2 (Zen BioScience,#R381559), anti-HO-1 (CST,#82551), anti-Caspase3 (CST,#14220), anti-cleaved-caspase3 (CST,#9664), anti-ERK (CST,#9102), anti-p-ERK (CST,#4370), anti-β-actin (Beyotime,#AF5003), anti-α-actin (CST,#19245) and anti-GAPDH (Beyotime,#AF1186), Anti-Rabbit secondary antibodies (Proteintech,#SA00001-2).

*Cell Cultivation*: FHS74Int cells were cultured in Dulbecco's modified Eagle's medium (DMEM) supplemented with 20% fetal bovine serum (FBS) and 1% penicillin/streptomycin (P/S). MCF10A cells were maintained in complete growth medium (Jennio-bio, #02202). The murine breast cancer cell line 4T1 was cultured in RPMI 1640 medium containing 10% FBS and 1% P/S. All cells were incubated under standard conditions (37°C, 5%$CO_2$, humidified atmosphere). Both normal cell lines (FHS74Int and MCF10A) were irradiated during the exponential growth phase with 6 Gy by either CONV-RT or FLASH-RT. Post-irradiation samples from normal cell lines were collected at various time points for subsequent cellular assays.

*NRF2 Knockdown in Cultured Cells*: For NRF2 knockdown, FHS74Int cells were seeded in 6-well plates one day in advance and allowed to reach approximately 70% confluence. Cells were transduced with lentiviral particles carrying either NRF2-targeting shRNA or non-targeting control shRNA (scrambled shRNA) in antibiotic-free medium. After 18 hours, the transduction mixture was replaced with fresh antibiotic-free medium and the cells were cultured for another 24 hours. Upon confirmation of stable transduction, the cells were selected with puromycin (2.5 μg/mL) for 6 days.

*Establishment of Xenograft Models*: As illustrated before, male BALB/C mice were used for 4T1 tumor model establishment. 4T1 suspensions containing $10^6$ 4T1 cells in 200μL of a 1:1 mixture of PBS and Matrigel (Corning, #356234) were subcutaneously injected into the thigh flank. Tumor volumes were continuously monitored until they exceeded 100 mm³, at which point various radiotherapy conditions were initiated. The tumor volume was calculated using calipers according to the formula, $0.5*width^2*length$.

*Measurement of ROS and MDA*: Cells were seeded in 96-well black plates and allowed to adhere until reaching approximately 70% confluency. Prior to irradiation, the cells were incubated with the fluorescent probe CM-H$_2$DCFDA (Thermo Fisher Scientific, C6827) at a working concentration of 5 μM for 30 min at 37°C. After irradiation, the cells were further incubated for 1 h under standard culture conditions. Fluorescence intensity was measured using a microplate reader (Excitation: 495 nm/Emission: 530 nm). Readings were taken both before and after irradiation to monitor temporal changes in ROS levels.

MDA content, an indicator of lipid peroxidation, was measured using a commercial assay kit (Dojindo, #M496) according to the manufacturer's instructions. The absorbance was measured at 532 nm using a microplate reader. MDA concentrations were determined based on a standard curve.

*Determination of GSH/GSSG and CAT:* Cells were collected 8h post-irradiation. Concentrations of GSH and GSSG were examined by the GSSG/GSH Quantification Kit II (Dojindo,#G263), and the GSH/GSSG ratio was calculated. For the CAT activity, 8h post-irradiation samples were lysed on ice and centrifuged to obtain clear supernatants. The reaction was initiated by adding hydrogen peroxide substrate, and the decrease in absorbance at 240 nm was monitored over 3 minutes. CAT activity was calculated based on the rate of H$_2$O$_2$ decomposition and normalized to total protein concentration.

*Simulation*: A simplified two-compartment model was used to study the chemical reactions and damage caused by FLASH irradiation, simulating interactions within the lipid and other molecular compartments. Reactants were categorized based on reaction types and rate constants. This model highlights the recombination of peroxyl radicals and their interactions with antioxidants across both compartments. The proportion of recombined peroxyl radicals was calculated to investigate the significance of recombination. The detailed computational procedures employed in this study were based on a previous study[20], which included the concentration of ROOH and the rates of radical recombination.

*Statistical analysis*: Data are presented as mean ± SD. One-way analysis of variance (ANOVA) was used to assess statistical significance among multiple groups, and the Student's t test was employed to determine the significance between each pair of experimental groups, with a significance level set at p < 0.05. All analyses were performed in GraphPad Prism 10.1.2.

**Conflict of Interest**

The authors declare no conflict of interest.

**Author Contributions**

Y.Z. and C.H. contributed equally to this work. Q.F. and T.H.contributed to the study conceptualization and design. Y.Z. and C.H. performed the animal experiments and data analysis. A.H. and W.Z. conducted the theoretical modeling and simulation. Y.W. and Y.Z. contributed to data curation. J.Q., J.W.,H.Z. assisted with FLASH irradiation and dosimetry measurements. and provided critical experimental conditions and resources, including the irradiation facility. Q.F., T.H. and H.Z. supervised the study. Q.F., T.H.,H.Z. and J.L. acquired funding. Q.F. and T.H wrote the orginal draft of the manuscript. Q.F., T.H., H.Z., W.W., and J.L. edited the manuscript. All authors reviewed and approved the manuscript.

**Data Availability Statement**

The data that support the findings of this study are openly available in NCBI under accession numbers GSE306333.

**Supporting Information**

Supporting Information is available from the Wiley Online Library or from the author.

**Unraveling the Redox Mechanisms Underlying FLASH Radiotherapy: Critical Dose Thresholds and NRF2-Driven Tissue Sparing**

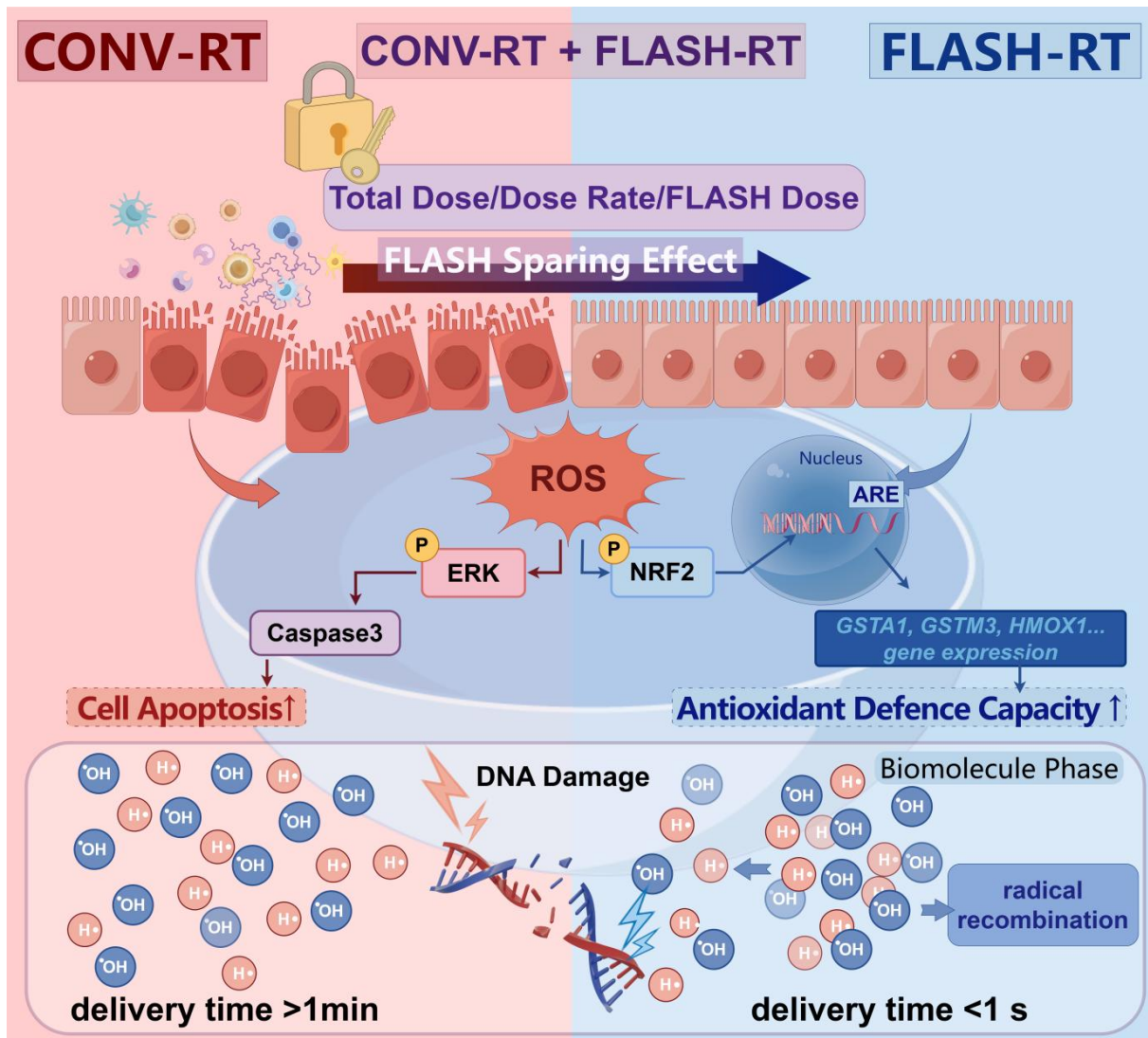

This study demonstrates that FLASH Radiotherapy maintains tumor control equivalent to CONV Radiotherapy while reducing intestinal toxicity through radical recombination, nuclear factor E2-related factor 2 (NRF2) activation, and extracellular regulated protein kinases (ERK) inhibition, even at substantially reduced doses (⩾ 6 Gy) provided a total dose threshold (15 Gy) is met, supporting redox-based strategies for clinical translation.

# Supporting Information

**Unraveling the Redox Mechanisms Underlying FLASH Radiotherapy: Critical Dose Thresholds and NRF2-Driven Tissue Sparing**

Yan Zhang 1†, Chenyang Huang 1†, Ankang Hu2,5, Yucheng Wang1, Yixun Zhu1, Wanyi Zhou 2,5, Jiaqi Qiu6, Jian Wang6, Qibin Fu1,4* Tuchen Huang 1,3,4* Hao Zha 2*, Wei Wang1,3,4, Junli Li2,5

**Table. S1** The scoring criteria of the histological score

| | submucosa inflammation | crypts structure | crypts regeneration | inflammation cells infiltration in lamina propria | thickening of the muscularis mucosae | epithelial damage |
|---|---|---|---|---|---|---|
| 0 | no active inflammation and erosions | no significant abnormalities | almost 100% being regenerative crypts | no inflammatory cell infiltration in the lamina propria | no thickening phenomenon | no Paneth cell metaplasia, goblet cell count is normal with no reduction |
| 1 | increased lymphocytes, plasma cells, and eosinophils | crypts are slightly deformed, with a reduction in numbers | 80-90% are regenerative crypts | a slight increase in lymphocytes and plasma cells | slight thickening | slight or localized Paneth cell metaplasia |
| 2 | a significant neutrophil infiltration | some crypts show deformation | the number of regenerative crypts begins to decrease significantly. | moderate infiltration of inflammatory cells in the lamina propria, such as neutrophils and eosinophils. | moderate thickening, with obvious fibrosis and muscularization. | Paneth cell metaplasia is evident, affecting multiple areas |
| 3 | The degree of infiltration further increases | crypt structure is markedly altered | 80% of crypts are deformed and have lost regenerative capacity. | severe infiltration of inflammatory cells in the lamina propria | further thickening | Paneth cell metaplasia is widely distributed. |
| 4 | extensive mucosal damage and inflammatory cell infiltration | Crypts are completely deformed and atrophic. | Almost no regenerative crypts within the section range. | Rare neutrophil infiltration of crypt epithelial cells, with further deepening of the degree of inflammatory cell infiltration. | Severe thickening. | Massive Paneth cell metaplasia occurs, with a reduction of goblet cells at the tips of the villi, and the epithelial layer may be accompanied by extensive erosion or ulceration. |

**Table. S2** Primer sequences of genes for RT-qPCR.

| Gene | Primer sequences |
|---|---|
| *β-actin* | Forward, 5'- GTGACGTTGACATCCGTAAAGA -3'<br>Reverse, 5'- GTAACAGTCCGCCTAGAAGCAC -3' |
| *Gsta1* | Forward, 5'- GGAATTTGATGTTTGACCAAGTGC -3'<br>Reverse, 5'- AAGGCAGGCAAGTAACGGTTTT -3' |
| *Hmox1* | Forward, 5'- AGCTGGTGATGGCTTCCTTGTAC -3'<br>Reverse, 5'- CTCCTCAGGGAAGTAGAGTGGG -3' |

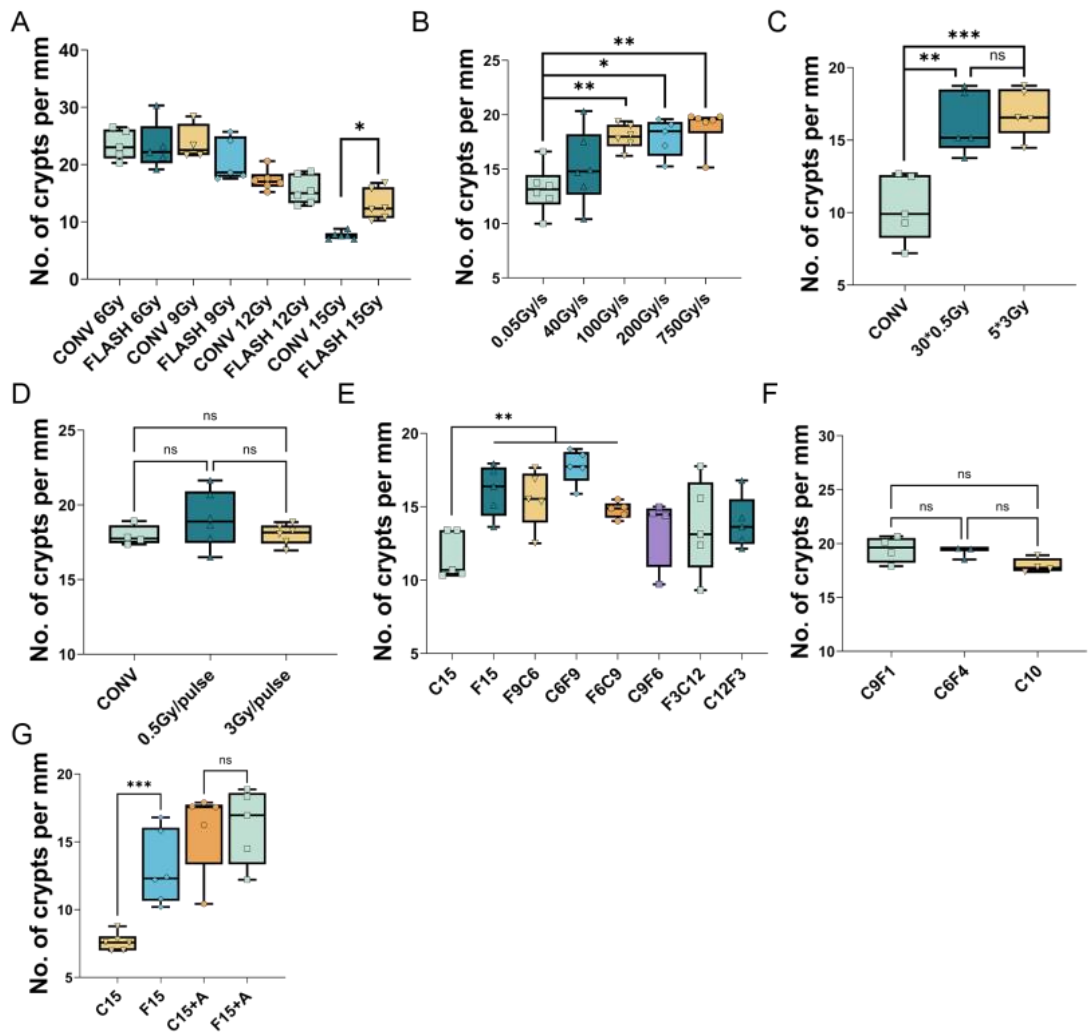

**Figure S1**. Quantification of number of crypts per millimeter in intestinal tissues after various radiation conditions. (A) At the condition of FLASH-RT (900Gy/s) or CONV-RT across dose escalation (n=6 per group, one-way ANOVA test). (B) At dose rate gradient at 15Gy. (C) Comparison with two pulse structure at 15 Gy and 150 Gy/s. (D) Comparison with two pulse structure at 10Gy and 150Gy/s. Comparison of different kinds of combined FLASH-RT/CONV-RT irradiation protocol at a dose of 15Gy (E) and 10Gy(F). (G) Combination of irradiation (CONV-RT or FLASH-RT) and amifostine injection.

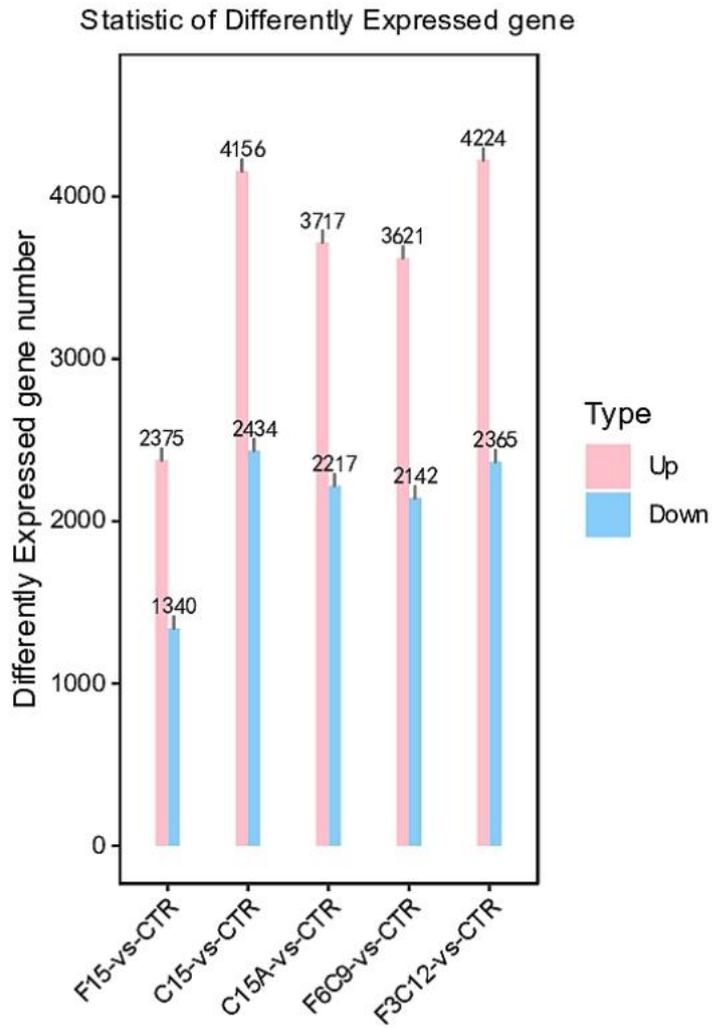

**Figure S2.** DEGs of the irradiated group (F15, C15, C15+A, F6C9, F3C12) compared to the CTR group.

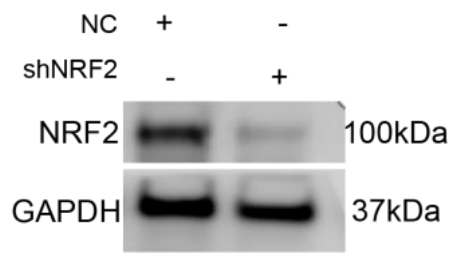

**Figure S3.** NRF2-knockdown efficiency validation in FHS74Int cells. (NC: non-treated cell; shNRF2: cells treated with an NRF2-silencing shRNA).